\documentclass{amsart}
\usepackage{amsmath, amssymb, mathtools}
\usepackage{tensor}
\usepackage{natbib}
\usepackage{booktabs}
\usepackage{url}
\usepackage{hyperref}
\newcommand*{\doi}[1]{DOI \href{https://doi.org/#1}{\texttt{#1}}}
\makeatletter
\def\bibinfo#1{%
  \@ifundefined{bibinfo@X@#1}%
    {\@firstofone}
    {\csname bibinfo@X@#1\endcsname}}
\makeatother
\hypersetup{final}

\usepackage[]{color}
\usepackage[obeyFinal]{todonotes}

\usepackage{array}
\usepackage{multirow}
\usepackage{tikz}
\usetikzlibrary{calc,decorations.pathmorphing}
\tikzset{
  treenode/.style={circle, draw=black,minimum size=0.3em,inner sep=0pt,fill},
  hybnode/.style={rectangle, draw=black,minimum width=0.3em, minimum height=0.3em,inner sep=0pt,fill},
  path/.style={->,decorate,decoration={snake,amplitude=.4mm,segment length=2mm,pre length=1mm,post length=1mm}}
}
 
\newcommand{\BTC}{\mathcal{BTC}}
\DeclareMathOperator{\THP}{TH}
\newcommand{\OTHP}{\overline\THP}
\DeclareMathOperator{\indeg}{indeg}
\DeclareMathOperator{\outdeg}{outdeg}
\newtheorem{thm}{Theorem}
\newtheorem{lem}[thm]{Lemma}
\newtheorem{prop}[thm]{Proposition}
\newtheorem{cor}[thm]{Corollary}
\theoremstyle{definition}
\newtheorem{defn}{Definition}
\newtheorem{rmk}{Remark}
\newenvironment{pf}{\begin{proof}}{\def\qed{}\end{proof}}

\title{Generation of Tree-Child phylogenetic networks}

\author{Gabriel Cardona}
\author{Joan Carles Pons}
\thanks{Partially supported by the Spanish Ministry of Economy and Competitiveness and European Regional Development Fund (MINECO/FEDER) project DPI2015-67082-P }
\address{Department of Mathematics and Computer Science, University of the Balearic Islands, Ctra. de Valldemossa, km. 7.5, E-07122 Palma, Spain}

\author{Celine Scornavacca}
\address{Institut des Sciences de l'\'Evolution,  Facult\'e des Sciences de Montpellier, Universit\'e Montpellier, 173567 Montpellier, France.}

\begin{document}

\begin{abstract}
  Phylogenetic networks generalize phylogenetic trees by allowing the modelization of events of reticulate evolution. Among the different kinds of phylogenetic networks that have been proposed in the literature, the subclass of binary tree-child networks is one of the most studied ones. However, very little is known about the combinatorial structure of these networks.

  In this paper we address the problem of generating  all possible binary tree-child networks with a given number of leaves in an efficient way via reduction/augmentation operations that extend and generalize analogous operations for phylogenetic trees and are biologically relevant. Since our solution is recursive, this also provides us with a recurrence relation giving an upper bound on the number of such networks.
  \end{abstract}

\maketitle

\section{Introduction}
Networks have been introduced in  phylogenetics to generalize the tree paradigm, which permits to  
 represent only descent with modification (i.e. speciation events). Phylogenetic networks allow to model evolutionary scenarios including a larger class of evolutionary events, such as  recombinations,  lateral gene transfers and hybridizations. 

In this paper, we shall focus on directed phylogenetic networks \citep[see][for a short survey on the phylogenetic network paradigm also covering undirected phylogenetic networks]{survey}. Mathematically, such networks are, in the broadest sense, directed acyclic graphs with a single node with no incoming edges --the \emph{root}-- representing the common ancestor of all the Operational Taxonomic Units (OTUs for short) under study, which are represented by the  nodes with no outgoing edges -- the \emph{leaves}-- of the graph; internal nodes represent either (hypothetical) speciations or (hypothetical) reticulated events. 
Unfortunately, this definition is too broad, both for representing biologically-meaningful evolutionary scenarios, and for giving objects that can be efficiently handled. 

So far, several restrictions on this general definition have been introduced in the literature. Some of them are based on biological considerations, while others have been introduced to artificially narrow the space of networks under study. 
This led to the introduction of a panoply of different classes of phylogenetic networks, such as time-consistent networks \citep{baroni2006}, regular networks \citep{Baroni2005}, orchard networks \citep{orchard}, galled trees \citep{galled_trees} and galled networks \citep{galled_net}, level-$k$ networks \citep{vanIersel2011}, tree-sibling networks \citep{Cardona2009}, tree-based networks \citep{Francis2015} and LGT networks \citep{cardona2015reconstruction}, just to name a few.

In this paper, we shall focus on binary tree-child networks (BTC networks, for short), which were introduced by \cite{Cardona2009} and are one of the most studied classes of phylogenetic networks \citep{VANIERSEL2010, vanIersel2014, Semple2016, Bordewich2018}. Mathematically, being tree-child means that every internal node is compelled to have at least a child node with a single incoming arc. Biologically, it boils down to say that every non-extant OTU has at least a direct descendant through mutation.


 The combinatorial study of phylogenetic networks is nowadays a challenging and  active field of research. Nevertheless, the problem of counting how many phylogenetic networks are in a given subclass  of networks is still open even for long-established classes. More precisely, this problem has been only recently solved for galled networks \citep{COuntGalled}; for other classes, including tree-child networks, we only have  
 asymptotic results \citep{McDiarmid2015,CountTC}. 
Associated to the problem of counting networks, we find the problem of their ``injective'' generation, 
 i.e. without having to check for isomorphism between pairs of constructed networks. 
 
 The main result of this paper is a systematic way of recursively generating, with unicity, all BTC networks with a given number of leaves. This generation relies on a pair of reduction/augmentation operations --both producing BTC networks-- where reductions decrease by one the number of leaves in a network, and augmentations increase it. The idea of using pairs of operation has already been used to deal either with other classes of phylogenetic networks \citep{cardona2008distance, cardona2009metrics2}, or for BTC networks but without the unicity feature \citep{orchard}.
 
 As interesting side product, this procedure gives a recursive formula providing an upper bound on the number of BTC networks. 
 
 The paper is organized as follows. In Section~\ref{sec:preliminaries}, we review the basic definitions that will be used throughout the paper. Section~\ref{sec:reduction} is devoted to the reduction procedure, while  in Section~\ref{sec:generation-networks} we introduce the augmentation operation and prove that any BTC network can be obtained, in an unique way, via a sequence of augmentation operations applied to the trivial network with one leaf. In Section~\ref{sec:bounding-offspring}, we show how to relax the conditions for the applicability of the augmentation operation  to obtain a recursive formula providing an upper bound on the number of BTC networks. In Section~\ref{sec:computations} we introduce the implementation of the algorithms presented in the paper, and some experimental results, including the exhaustive generation of all BTC networks with up to six leaves and an upper bound of their number up to ten leaves.
  Finally, in Section~\ref{sec:conclusions} we discuss how our reduction/augmentation operations extend and generalize analogous operations for phylogenetic trees.


\section{Preliminaries}
\label{sec:preliminaries}

Throughout this paper, a \emph{tree node} in a directed graph is a node $u$ whose pair of degrees $d(u)=(\indeg u,\outdeg u)$ is $(1,0)$ for the \emph{leaves}, $(0,2)$ for the \emph{roots}, or $(1,2)$ for \emph{internal} tree nodes; a \emph{hybrid node} is a node $u$ with $d(u)=(2,1)$.

A \emph{binary phylogenetic network} over a set $X$ of taxa is a directed acyclic graph with a single root such that all its nodes are either tree nodes or hybrid nodes, and whose leaf set is bijectively labeled by the set $X$. In the following, we will implicitly identify every leaf with its label.
A binary phylogenetic network is \emph{tree-child} if every node either is a leaf or has at least one child that is a tree node \citep{Cardona2009}; in particular, the single child of a hybrid node must be a tree node.
We will denote by $\BTC_n$ the set of binary tree-child phylogenetic networks over the set $[n]=\{1,\dots,n\}$.


An \emph{elementary node} in a directed graph is a node $u$ with $d(u)=(1,1)$ or $d(u)=(0,1)$. An \emph{elementary path} $p$ is a path $u_1,\dots,u_k$ composed of  elementary nodes such that neither the single parent of $u_1$ (if it exists) nor the single child of $u_k$ are elementary. We  call these last two nodes respectively the \emph{grantor} (if this node is well-defined) and \emph{heir} of the nodes in the elementary path. In case of an elementary node, its grantor and heir are those of the nodes in the single elementary path that contains the given node. The \emph{elimination} of an elementary path $p$ consists in deleting all nodes in $p$, together with their incident arcs, and adding an arc between the grantor and the heir of $p$ (provided that the grantor exists; otherwise, no arc is added). The elimination of an elementary node is defined as the elimination of the elementary path that contains the given node.

Given a node $u$, we can \emph{split} it by adding a new node $\tilde u$, an arc $(\tilde u,u)$, and replacing every arc $(v,u)$ with $(v,\tilde u)$. If $u$ is a tree node, then $\tilde u$ is an elementary node whose heir is $u$, and the elimination of $\tilde u$ recovers the original network. The successive splitting (say $k$ times) of a tree node $u$ generates an elementary path formed by $k$ nodes, whose heir is $u$, and whose elimination recovers the original network.

\section{Reduction of networks}
\label{sec:reduction}

The goal of this section is to define a reduction procedure on BTC networks that can be applied to any such network, and producing a  BTC network with one leaf less. By successive application of this procedure, any BTC network can thus be reduced to the trivial network with a single leaf.

We start by associating to each leaf $\ell$ a path whose removal will produce the desired reduction (up to elementary paths).

Let $\ell$ be a leaf of a BTC network $N$. A \emph{pre-TH-path} for $\ell$ is a path $u_1,\dots,u_r=\ell$ such that: 
\begin{enumerate}
    \item Each node $u_i$ in the path is a tree node.
    \item For each $i=1,\dots,r-1$, the child of $u_i$ different from $u_{i+1}$, denoted by $v_{i}$, is a hybrid node.
    \item For each $i\neq j$, we have that $v_i\neq v_j$.
\end{enumerate}
A \emph{TH-path} is a maximal pre-TH-path, i.e. a pre-TH-path that cannot be further extended. Note that, since all nodes in a pre-TH-path $p$ are tree nodes, if $p$ can be extended by prepending one node, then this extension is unique. Hence, starting with the trivial pre-TH-path formed by the leaf $\ell$ alone, and extending it by prepending the parent of the first node in the path as many times as possible, we obtain a TH-path that is unique by construction.
Let $u_1,\dots,u_r=\ell$ be a TH-path; different possibilities may arise that make it maximal: (1) $u_1$ is the root of $N$; (2) the parent of $u_1$, call it $x$, is a hybrid node; (3) $x$ is a tree node whose both children are tree nodes; (4) $x$ is a parent of $v_i$ for some $i \in [r-1]$. (We shall see in Lemma \ref{lemma:noRoot} that the first case  cannot hold).

For each leaf $\ell$, we denote by $\THP(\ell)$ its single TH-path and by $\THP(\ell)_1$ the first node of this path.
Note that we allow the case $r=1$. 
In this case, if we are not in a trivial BTC network (i.e.\ a network consisting of a single node),  the parent of $\ell$ is either a hybrid node, or a tree node whose two children are tree nodes.




\begin{lem}\label{lemma:noRoot}
  Let $N$ be a non-trivial BTC network  and let $\ell$ be any of its leaves. Then, $\THP(\ell)_1$ cannot be the root of $N$.
\end{lem}

\begin{pf}
  Let $u_1,\dots,u_r=\ell$ be the path $\THP(\ell)$ and assume for the sake of contradiction that $u_1$ is the root of $N$. For each $i=1,\dots,r-1$, let $v_i$ be the hybrid node that is a child of $u_i$ and $x_i$ the parent of $v_i$ different from $u_i$;  recall that $x_i$ does not belong to $\THP(\ell)$ by the definition of a pre-TH-path. Since $u_1$ is the root of $N$, every node of $N$ either belongs to the path $\THP(\ell)$ or is descendant of a node in $\{v_i\mid i\in[r-1]\}$.
  In particular, for each $i\in[r-1]$, there exists some $\sigma(i)\in[r-1]$ such that
  $x_i$ is descendant of $v_{\sigma(i)}$, and since this node is descendant of $x_{\sigma(i)}$, $x_i$ is descendant of $x_{\sigma(i)}$. Hence, starting with $x_1$ we get a sequence
  $x_1, x_{\sigma(1)},  x_{\sigma(\sigma(1))}, \dots$
  where each node in the sequence is a descendant of the following one.
  Since there is a finite number of nodes, at some point we find a repeated node, which means that $N$ contains a cycle and hence we have a contradiction.
  \qed
\end{pf}

We say that a leaf $\ell$ is of \emph{type} $T$ (resp. of \emph{type} $H$) if the parent of $\THP(\ell)_1$ is a tree node (resp. a hybrid node). If $\ell$ is of type $H$, we  indicate by $\OTHP(\ell)$ the path obtained by prepending to $\THP(\ell)$ the parent of $\THP(\ell)_1$. 
For convenience, we let $\OTHP(\ell)=\THP(\ell)$ if $\ell$ is of type $T$.

\begin{figure}
  \centering
  \begin{tikzpicture}
    \node[treenode,label=left:$w_1$] (w) at (0,1) {};
    \node[treenode,label=left:{$u_0=u_1$}] (u1) at (0,0) {};
    \node[treenode,label=left:{$u_i$}] (u2) at (0,-1.5) {};
    \node[treenode,label=left:{$u_{r-1}$}] (u3) at (0,-3) {};
    \node[treenode, label=left:{$u_r=\ell$}] (u4) at (0,-4) {};
    \node[hybnode
    ] (v1) at ($(u1) +(-30:1)$) {};
    \node[hybnode,label=above:$v_i$] (v2) at ($(u2) +(-30:1)$) {};
    \node[hybnode
    ] (v3) at ($(u3) +(-30:1)$) {};
    \coordinate (v1p) at ($(v1) +(30:0.5)$);
    \coordinate (v1c) at ($(v1) +(-90:0.5)$);
    \draw[->] (v1p)--(v1);
    \draw[->] (v1)--(v1c);
    \draw[->,dashed] (v1p) to 
    (v1c);
    \coordinate[label=right:{$x_i$}] (v2p) at ($(v2) +(30:0.5)$);
    \coordinate[label=right:{$y_i$}] (v2c) at ($(v2) +(-90:0.5)$);
    \draw[->] (v2p)--(v2);
    \draw[->] (v2)--(v2c);
    \draw[->,dashed] (v2p) to 
    (v2c);
    \coordinate (v3p) at ($(v3) +(30:0.5)$);
    \coordinate (v3c) at ($(v3) +(-90:0.5)$);
    \draw[->] (v3p)--(v3);
    \draw[->] (v3)--(v3c);
    \draw[->,dashed] (v3p) to 
    (v3c);
    \coordinate[label=right:$t_1$] (wc) at ($(w) +(-30:0.5)$) {};
    \coordinate[label=right:$z_1$] (wp) at (0,1.5) ;
    \draw[->] (wp)--(w);\draw[->] (w)--(wc);
    \draw[->,dashed] (wp) to 
    (wc);
    \draw[->] (w)--(u1);
    \draw[path] (u1)--(u2); 
    \draw[path] (u2)--(u3); 
    \draw[->] (u3)--(u4);
    \draw[->] (u1)--(v1);
    \draw[->] (u2)--(v2);
    \draw[->] (u3)--(v3);
    \draw[dotted] ($(u4)+(-0.2,-0.2)$) rectangle ($(u1)+(0.2,0.2)$);
  \end{tikzpicture}
  \qquad
  \begin{tikzpicture}
    \node[hybnode,label=left:$u_0$] (u0) at (0,1) {};
    \node[treenode,label=left:{$u_1$}] (u1) at (0,0) {};
    \node[treenode,label=left:{$u_i$}] (u2) at (0,-1.5) {};
    \node[treenode,label=left:{$u_{r-1}$}] (u3) at (0,-3) {};
    \node[treenode, label=left:{$u_r=\ell$}] (u4) at (0,-4) {};
    \node[hybnode
    ] (v1) at ($(u1) +(-30:1)$) {};
    \node[hybnode,label=above:$v_i$] (v2) at ($(u2) +(-30:1)$) {};
    \node[hybnode
    ] (v3) at ($(u3) +(-30:1)$) {};
    \coordinate (v1p) at ($(v1) +(30:0.5)$);
    \coordinate (v1c) at ($(v1) +(-90:0.5)$);
    \draw[->] (v1p)--(v1);
    \draw[->] (v1)--(v1c);
    \draw[->,dashed] (v1p) to 
    (v1c);
    \coordinate[label=right:{$x_i$}] (v2p) at ($(v2) +(30:0.5)$);
    \coordinate[label=right:{$y_i$}] (v2c) at ($(v2) +(-90:0.5)$);
    \draw[->] (v2p)--(v2);
    \draw[->] (v2)--(v2c);
    \draw[->,dashed] (v2p) to 
    (v2c);
    \coordinate (v3p) at ($(v3) +(30:0.5)$);
    \coordinate (v3c) at ($(v3) +(-90:0.5)$);
    \draw[->] (v3p)--(v3);
    \draw[->] (v3)--(v3c);
    \draw[->,dashed] (v3p) to 
    (v3c);
    %
    %
    \draw[->] (w)--(u1);
    \draw[path] (u1)--(u2); 
    \draw[path] (u2)--(u3); 
    \draw[->] (u3)--(u4);
    \draw[->] (u1)--(v1);
    \draw[->] (u2)--(v2);
    \draw[->] (u3)--(v3);
    %
    \node[treenode,label=left:$w_2$] (w') at ($(u0) + (30:1)$) {};
    \node[treenode,label=right:$w_1$] (w) at ($(u0) + (150:1)$) {};
    \draw[->] (w')--(u0);
    \draw[->] (w)--(u0);
    %
    \coordinate[label=left:$z_1$] (wp) at ($(w) +(90:0.5)$);
    \coordinate[label=left:$t_1$] (wc) at ($(w) +(210:0.5)$);
    \draw[->] (wp)--(w);
    \draw[->] (w)--(wc);
    \draw[->,dashed] (wp) to 
    (wc);
    \coordinate[label=right:$z_2$] (w'p) at ($(w') +(90:0.5)$);
    \coordinate[label=right:$t_2$] (w'c) at ($(w') +(-30:0.5)$);
    \draw[->] (w'p)--(w');
    \draw[->] (w')--(w'c);
    \draw[->,dashed] (w'p) to 
    (w'c);
    %
    \draw[dotted] ($(u4)+(-0.2,-0.2)$) rectangle ($(u0)+(0.2,0.2)$);
  \end{tikzpicture}
  \caption{Sketches of 
  a $T$-reduction (left) and 
  a $H$-reduction (right). Tree nodes are represented by circles and hybrid nodes by squares.
  The nodes inside the dotted box form $\OTHP(\ell)$ and will be removed, which will create elementary nodes that will be substituted by the dashed arcs.}
  \label{fig:T-red}
\end{figure}
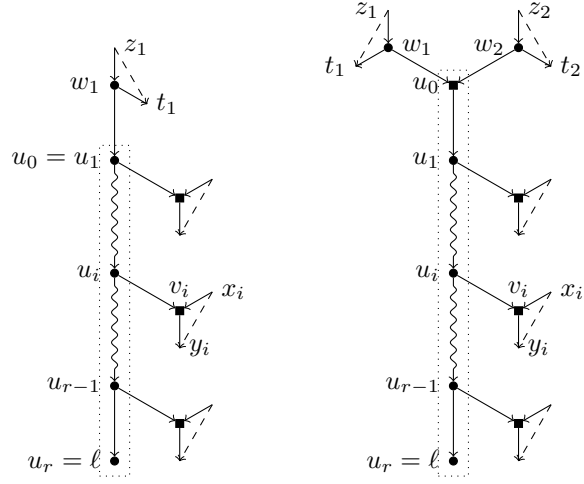

\begin{defn}
  \label{def:reductions}
Let $\ell$ be a leaf in a BTC network $N$. We define the \emph{reduction of $N$ with respect to $\ell$} as the result of the following procedure:
\begin{enumerate}
\item Delete all nodes in $\OTHP(\ell)$ (together with any arc incident on them).
\item Eliminate all elementary nodes.
\end{enumerate}
We indicate this reduction by $R(N,\ell)$. If we want to emphasize the type of the deleted leaf, we  indicate the reduction by $T(N,\ell)$ and say it is a $T$-reduction if $\ell$ is of type $T$, or by $H(N,\ell)$ and say that it is a $H$-reduction if $\ell$ is of type $H$. 
\end{defn}

To ease of reading, we shall introduce some  notations:
\begin{defn}
  \label{defn:elementarynodes}
Let $u_1,\dots,u_r=\ell$ be the path $\THP(\ell)$ and let $u_0$ be the first node in $\OTHP(\ell)$. For each $i \in [r-1]$, $v_i$ is the hybrid child of $u_i$, $x_i$ the parent of $v_i$ different from $u_i$, and $y_i$ the single child of $v_i$. The parent(s) of $u_0$ is $w_1$ (are $w_1,w_2$); the node $w_j$ is always a tree node, $z_j$ is its parent (if it exists, since $w_j$ could be the root of $N$), and $t_j$ its child different from $u_0$. 
\end{defn}
In the following, we shall use the notations given in the definition above, which are illustrated in Figure~\ref{fig:T-red}.

\begin{rmk}
  \label{rmk:elementarynodes}
  Since $N$ is tree-child, the nodes $y_i$ are always tree nodes, and so are $t_1$ and $t_2$ in case of an $H$-reduction. In case of a $T$-reduction, by definition of a TH-path, $t_1$ is either a tree node or coincides with one  of the hybrid nodes $v_i$.
Also, the removal of the arcs of the form $(u_i,v_i)$  and $(w_j,u_0)$ makes nodes $v_i$ and  $w_j$ elementary in $N\setminus \OTHP(\ell)$, where $i \in [r-1]$, and $j=1 $ for $T$-reductions and $j \in [2]$ for $H$-reductions. Since no other arc is removed, no other node can be elementary. In order to find the heirs of nodes $v_i$ and  $w_j$, we must analyse under which circumstances two of these elementary nodes are adjacent in $N\setminus \OTHP(\ell)$. 
\begin{enumerate}
    \item If we had that two nodes $v_i$ and $v_j$ were connected by an arc in $N\setminus \OTHP(\ell)$, then the single child of a hybrid node in $N$ would be also a hybrid. This contradicts the fact  that $N$ is tree-child.
    \item The existence of an arc $(v_i,w_j)$ would imply the existence of a cycle in $N$, which is impossible. 
    \item Consider now the case of an arc $(w_j,v_i)$. In case of an $H$-reduction, it would imply that  both children of $w_j$ are hybrid nodes, which is impossible. However, such an arc can be present in a $T$-reduction: when $t_1$ is equal to $v_i$. In this last case, $w_1$ and $v_i$ form an elementary path in $N\setminus \OTHP(\ell)$ and their common 
      heir is $y_i$.
    \item Finally, in case of an $H$-reduction, it can exist an arc between $w_1$ and $w_2$, say that the arc is $(w_1,w_2)$ (which implies, $t_1=w_2$, $z_2=w_1$). In this case, $w_1$ and $w_2$ form an elementary path in $N\setminus \OTHP(\ell)$ and their common
      heir is $t_2$.
\end{enumerate}
In all other cases, the elementary nodes $v_i$ and $w_j$ are isolated, and their respective heirs are $y_i$ and $t_j$.
\end{rmk}

We study now what we call the \emph{recovering data} of a reduction. This information will be used in the next section to recover the original network from its reduction.

\begin{defn}
  The \emph{recovering data} of the reduction $N'=R(N,\ell)$ is the pair $(S_1,S_2)$, where: 
  \begin{itemize}
      \item $S_1$ is the multiset of the nodes of $N'$ that are heirs of the nodes $w_j$. 
      The cardinality of $S_1$ (as a multiset) is either $1$ or $2$, depending on the type of the reduction, and will be denoted by $|S_1|$.
      \item $S_2$ is the tuple $(y_1,\dots,y_{r-1})$ of nodes of $N'$, which are the heirs of the nodes $v_i$. This tuple could be empty, corresponding to the case $r=1$.
  \end{itemize}
 \end{defn}

We introduce now a set of conditions on multisets and tuples of nodes, and prove that the recovering data associated to any of the defined reductions satisfies them.

\begin{defn}\label{dfn:feasible}
  Given a BTC network $N'$ and a pair $(S_1,S_2)$ with
    \begin{itemize}
      \item $S_1$ a multiset of tree nodes of $N'$,
      \item $S_2=(y_1,\dots,y_{r-1})$, with $r\ge1$, an (eventually empty) tuple of $r-1$ tree nodes of $N'$,
  \end{itemize}
  consider the following set of conditions:
  \begin{enumerate}
      \item For every $i,j\in [r-1]$ with $i\neq j$, the nodes $y_i$ and $y_j$ are different, and if they are siblings, then $y_i\in S_1$ or $y_j\in S_1$.
      \item For every $i\in[r-1]$, if $y_i$ is the child of a hybrid node or has a hybrid sibling, then $y_i\in S_1$.
      \item No node in $S_1$ is a proper descendant of any node in $S_2$.
      \item[4T.] $|S_1|=1$.
      \item[4H.]  $|S_2|=2$ and no node of $S_1$ appears in $S_2$.
  \end{enumerate}
  We say that $(S_1,S_2)$ is \emph{$T$-feasible} if it satisfies conditions 1, 2, 3, and 4T, and \emph{$H$-feasible} if it satisfies conditions  1, 2, 3, and 4H. Finally,  we say that $(S_1,S_2)$ is \emph{feasible} if it is either $T$-feasible or $H$-feasible.
\end{defn}




\begin{prop}
  Let $N'=T(N,\ell)$ be a $T$-reduction of a BTC network $N$. Then,  its recovering data  $(\{\tau_1\},(y_1,\dots,y_{r-1}))$ is $T$-feasible.
\end{prop}

\begin{pf}
First, note that, by Remark \ref{rmk:elementarynodes}, all nodes in  $(\{\tau_1\},(y_1,\dots,y_{r-1}))$ are tree nodes and  that Condition 4T holds trivially. 
Note also that $\tau_1$ is equal to $y_i$ if $t_1=v_i$, or  to $t_1$ if this node is different from all the nodes $v_i$.
We now prove that Conditions 1, 2 and 3 hold: 
\begin{enumerate}
    \item If $y_i=y_j$, then in $N$ we have $v_i=v_j$, which is impossible by definition of TH-path. If $y_i$ and $y_j$ are siblings in $N'$ but none of these nodes is equal to $\tau_1$, then $v_i$ and $v_j$ are siblings in $N$, which implies that their common parent has two hybrid children, which is impossible in a BTC network.
    \item If $y_i$ is the child in $N'$ of a hybrid node and $\tau_1\neq y_i$, then in $N$ we have that $v_i$, which is a hybrid node, is the child of a hybrid node, which is impossible in a tree-child network. Analogously, if $y_i$ has a sibling in $N'$ which is a hybrid node, and $y_i\neq \tau_1$, then in $N$ we have that $v_i$ is sibling of another hybrid node, which is again impossible.
    \item The existence of  a non-trivial path in $N'$ from $y_i$ to $\tau_1$ would, by construction, imply the existence of a path from $y_i$ to $w_1$ in $N$. Since there exists also a path in $N$ from $w_1$ to $y_i$, this would contradict the fact that $N$ is a DAG.
    \qed
\end{enumerate}
\end{pf}

\begin{prop}
  Let $N'=H(N,\ell)$ be an $H$-reduction of a BTC network $N$. Then,  its recovering data $(\{\tau_1,\tau_2\},(y_1,\dots,y_{r-1}))$ is $H$-feasible.
\end{prop}

\begin{pf}
Again we have, by Remark \ref{rmk:elementarynodes}, that all nodes in the recovering data are tree nodes. Additionally, by the same remark, we have that $|S_1|=2$ --and hence the first part of Condition 4H holds-- and if $(w_1,w_2)$ is an arc of $N$, then $S_1=\{t_2,t_2\}$, otherwise $S_1=\{t_1,t_2\}$ with $t_1\neq t_2$.
Note that Condition $3$ implies that Conditions 1 and 2 can be simplified as follows: 
for all $i,j\in [r-1]$ with $i\neq j$, $y_i$ and $y_j$ are neither equal nor siblings, and for all $i\in[r-1]$, $y_i$ is neither the child nor the sibling of a hybrid node.
  
Conditions 1 and Conditions 2 and 3 in their simplified form follow using the same arguments as in the previous proposition. As for the condition 4H,
the nodes $\tau_1$ and $\tau_2$ are different from the nodes $y_i$ since the parents of $\tau_1$ and $\tau_2$ in $N$ are tree nodes, while the parent of each of the nodes $y_i$ is hybrid.
    \qed
\end{pf}


The following proposition is the main result of this section, since it shows that the reduction that we have defined, when applied to a BTC network, gives another BTC network with one leaf less. Hence,  successive applications of these reductions reduce any BTC network to the trivial BTC network.

\begin{prop}
  Let $N$ be a BTC network over $X$ and $\ell$ one of its leaves. Then, $R(N,\ell)$ is a BTC network over $X\setminus\{\ell\}$.
\end{prop}

\begin{pf}
First, it is easy to see that, since no new path is added, the resulting directed graph is still acyclic.

  Then, we need to check that  $R(N,\ell)$ is binary. To do so, we start noting that every node in $N\setminus\OTHP(\ell)$ is either a tree node, a hybrid node, or an elementary node. Indeed, the removal of $\OTHP(\ell)$ (Phase 1 of Definition \ref{def:reductions}) only affects the nodes adjacent to this path, that is the nodes $v_i$ and $w_i$,  which, as shown in Remark~\ref{rmk:elementarynodes}, become elementary. The elimination of all elementary nodes (Phase 2 of Definition \ref{def:reductions}) does not affect the indegree and outdegree of any other node, apart when the root $\rho$ of $N\setminus\OTHP(\ell)$ is elementary. In such a case, the heir of $\rho$ becomes the new root. Hence, $R(N,\ell)$  is binary and rooted.

  Note also that the set of leaves of $R(N,\ell)$ is $X\setminus\{\ell\}$, since in $N\setminus\OTHP(\ell)$ no node becomes a leaf and the only leaf that is removed is $\ell$.
  
  Finally, we need to prove that $R(N,\ell)$ is tree-child. Note that, from what we have just said about how the reduction affects indegrees and outdegrees of the nodes that persist in the network, it follows that each hybrid node of $R(N,\ell)$ is also a hybrid node of $N$, and that its parents in $R(N,\ell)$ are the same as in $N$. It follows that no node in $R(N,\ell)$ can have that all its children are hybrid, since this would imply that $N$ is not tree-child, a contradiction.
  \qed
\end{pf}


\begin{cor}\label{cor:decomposition-of-network}
  Let $N\in\BTC_n$ be a BTC network over $[n]$. Let $N_n=N$ and define recursively $N_{i}=R(N_{i+1},i+1)$ for each $i=n-1,n-2,\dots,1$. Then, $N_i$ is a BTC network over $[i]$. In particular, $N_{1}$ is the trivial BTC network with its single node labeled by $1$.
\end{cor} 

We finish this section with the computation of the number of tree nodes and hybrid nodes that the reduced network has, both in terms of the original network and of the reduction operation that has been applied. But before, we give an absolute bound on the number of these nodes in terms of the number of leaves. 

\begin{lem}\label{lem:number_nodes}
Let $N$ be BTC network over $[n]$ with $t$ tree nodes and $h$ hybrid nodes. Then $t-h=2n-1$, $h\le n-1$ and $t\le 3n-2$.
\end{lem}

\begin{pf}
%
The equality $t-h=2n-1$ follows easily from the handshake lemma taking into account the number of roots, internal tree nodes, leaves and hybrid nodes in $N$, and their respective indegrees and outdegrees. The inequality  $h\le n-1$ is shown in Proposition~1 in \citep{Cardona2009}, and the last inequality is a simple consequence of the equality and the inequality already proved.
  \qed
\end{pf}

\begin{prop}
  Let $N$ be a BTC network and $\ell$ one of its leaves, and $N'=R(N,\ell)$. Let $t,h$ (resp. $t',h'$) the number of tree nodes and hybrid nodes of $N$ (resp. of $N'$). Then
  $$t'=t-|\OTHP(\ell)|-1,\qquad h'=h-|\OTHP(\ell)|+1,$$
  where $|\OTHP(\ell)|$ is the number of nodes in $\OTHP(\ell)$.
\end{prop}

\begin{pf}
  Since the number of tree nodes and hybrid nodes are linked by the equality in Lemma~\ref{lem:number_nodes}, it is enough to prove that $h'=h-|\OTHP(\ell)|+1$. From the discussion in Remark~\ref{rmk:elementarynodes}, it is straightforward to see that the number of hybrid nodes in $N$ that are not in $N'$ is $r-1$ if $\ell$ is of kind $T$, and $r$ otherwise. Hence, in both cases we have $h'=h-(|\OTHP(\ell)|-1)$ and the result follows. 
  \qed
\end{pf}

\section{Generation of networks}
\label{sec:generation-networks}

In this section, we consider the problem of how to revert the reductions defined in the previous section, taking as input the reduced network and its recovering data. This will allow us to define a procedure that, starting with the trivial BTC network with one leaf, generates all the BTC networks with any number of leaves in an unique way. 

We start by defining two augmentation procedures that take as input a BTC network and a feasible pair, and produce a BTC network with one leaf more.

\begin{defn}
\label{T-augmentation}
  Let $N$ be a BTC network over $X$, $\ell$ a label not in $X$, and  $(\{\tau_1\},(y_1,\dots,y_{r-1}))$ a $T$-feasible pair.
  We apply the following operations to $N$:
  \begin{enumerate}
  \item Create a path of new nodes $u_1,\dots,u_r$.
  \item Split the node $\tau_1$ creating one elementary node $w_1$ and add an arc $(w_1, u_1)$.
  \item For each node $y_i$, split it introducing one elementary node $v_i$ and add an arc $(u_i,v_i)$.
  \item Label the node $u_r$  by $\ell$.
  \end{enumerate}
    We denote by $T^{-1}(N,\ell;\{\tau_1\},(y_1,\dots,y_{r-1}))$ the resulting network and say that it has been obtained by an \emph{augmentation operation of type} $T$.
\end{defn}

Note that the order in which steps 2 and 3 are done is relevant in the case that $\tau_1=y_i$ for some $i\in[r-1]$. In such a case, two  nodes $w_1$ and $v_i$ are created, linked by an arc $(w_1,v_i)$.

\begin{prop}\label{thm:correctness-T-augmentation}
  Using  the notations of Definition~\ref{T-augmentation}, the network 
  $$\tilde N=T^{-1}(N,\ell;\{\tau_1\},(y_1,\dots,y_{r-1}))$$ 
  is a  BTC network over $X\cup\{\ell\}$. Moreover, if $N$ has $h$ hybrid nodes, then $\tilde N$ has $h+r-1$ hybrid nodes.
\end{prop}

\begin{pf}
We first check that the resulting directed graph is acyclic. Let us assume that $\tilde N$ contains a cycle. If we define  $U_1=\{u_1,\dots,u_r\}$ and $U_2=V(\tilde N)\setminus U_1$, we have that the only arcs connecting $U_1$ with $U_2$ are $(u_i,v_i)$ (with $i=1,\dots,r-1$), and $(w_1,u_1)$ is the only arc connecting $U_2$ with $U_1$. The cycle can be contained neither inside $U_1$, since these nodes are linked by a single path, nor inside $U_2$, since otherwise $N$ would contain a cycle. Hence, the cycle must contain at least the arc $(w_1,u_1)$ and an arc $(u_i,v_i)$. This implies the existence of a path from $v_i$ to $w_1$ visiting only nodes in $U_2$, which in turn means that $N$ contains a path from $y_i$ to $\tau_1$, against Condition 3 of Definition  \ref{dfn:feasible}.

Note that the nodes in $U_1$ are tree nodes by construction. Also by construction, the node $w_1$ is a tree node,  the nodes $v_i$ are hybrid nodes and $u_r$ is a leaf which is labelled with $\ell$. Finally, the other nodes keep the same degrees they had in $N$ and hence $\tilde N$ is a  binary phylogenetic network  over $X\cup\{\ell\}$ with $h+r-1$ hybrid nodes.

Since $N$ is tree-child, in order to check that $\tilde N$ is also tree-child, we only need to check the newly added  hybrid nodes, which are the parents of the nodes $v_i$. 

Let us first consider the case that $\tau_1\neq y_i$ for all $i\in[r-1]$.
For each node $v_i$, its parents are $u_i$ and the parent $x_i$ of $y_i$ in $N$. The node $u_i$ is by construction a tree node whose other child is $u_{i+1}$, which, in turn, is a tree node. 
Since $\tau_1\neq y_i$, by Condition 2 of Definition \ref{dfn:feasible},  $y_i$ can have neither a hybrid parent nor a hybrid sibling, and it cannot be a sibling of any other node $y_j$ with $j\in[r-1]$. This latter restriction implies that  $y_i$ has the same sibling $\tilde x_i$ in $N$ and $\tilde N$. Thus both $x_i$ and $\tilde x_i$ are not hybrid nodes, and the network is tree-child.


Let us now consider the case that $\tau_1=y_i$ for a single choice of $i\in[r-1]$. The hybrid node $v_i$ in $\tilde N$ has as parents the nodes $w_1$ and $u_i$, and these two nodes have as respective children $u_1$ and $u_{i+1}$, which are tree nodes. For each other node $v_j$ with $j\neq i$ and such that $y_j$ is a not sibling of  $y_i$, the same argument as in the previous case proves that both parents of $v_j$ have a tree child. If $y_j$ is a sibling of $y_i$, it is easy to see that the parent of $v_j$ is still tree-child since it has $w_1$ as child.
\qed
\end{pf}

\begin{defn}\label{dfn:H-augmentation}
  Let $N$ be a BTC network over $X$, $\ell$ a label not in $X$, and  $(\{\tau_1,\tau_2\},(y_1,\dots,y_{r-1})$  a $H$-feasible pair. 
  We apply the following operations to $N$:
  \begin{enumerate}
  \item Create a path of new nodes $u_0,u_1,\dots,u_r$.
  \item Split each of the nodes $\tau_i$ introducing one elementary node $w_i$ and add an arc from $w_i$ to $u_0$. Note that, if $\tau_1=\tau_2$, two consecutive elementary nodes must be created.
  \item For each node $y_i$, split it introducing one elementary node $v_i$ and add an arc $(u_i,v_i)$.
  \item Label the node $u_r$  by $\ell$.
  \end{enumerate}
We  denote by $H^{-1}(N,\ell;\{\tau_1,\tau_2\},(y_1,\dots,y_{r-1}))$ the resulting network and say that it has been obtained by an augmentation operation of type $H$.
\end{defn}

\begin{prop}
  Using  the notations of Definition~\ref{dfn:H-augmentation}, the network
  $$\tilde N=H^{-1}(N,\ell;\{\tau_1,\tau_2\},(y_1,\dots,y_{r-1}))$$ 
  is a BTC network over $X\cup\{\ell\}$. If $N$ has $h$ hybrid nodes, then $\tilde N$ has $h+r$ hybrid nodes.
\end{prop}

\begin{pf}
  The proof is completely analogous to that of Proposition~\ref{thm:correctness-T-augmentation}, taking into account that one extra hybrid node is created.
\qed
\end{pf}

Given a BTC network over $X$, a label $\ell\notin X$ and a feasible pair $(S_1,S_2)$, in order to unify notations  we define the augmented network $R^{-1}(N,\ell,S_1,S_2)$ as $T^{-1}(N,\ell,S_1,S_2)$, if $|S_1|=1$, and as $H^{-1}(N,\ell,S_1,S_2)$, if $|S_1|=2$. Also, we shall generically say that the \emph{offspring} of a BTC network is the set of networks that can be obtained from it by means of augmentation operations.

Our next goal is to prove that different augmentation operations applied to a same BTC network or different BTC networks over the same set of taxa  provide different networks. We start with the case of different networks.

\begin{prop}\label{prp:different_nets}
      Let $\tilde N_1$ and $\tilde N_2$ be two BTC networks, both obtained by one augmentation operation applied to two non-isomorphic BTC networks $N_1$ and $N_2$ over the same set of taxa $X$. Then $\tilde N_1$ and $\tilde N_2$ are not isomorphic.
\end{prop}

\begin{pf}
    If $\tilde N_1$ and $\tilde N_2$ have different set of labels, then it is clear that they are not isomorphic. We can therefore assume that both augmentation operations introduced the same new leaf $\ell$.
    Suppose that $\tilde N_1\simeq \tilde N_2$. Then $R(\tilde N_1,\ell)\simeq R(N_2,\ell)$. Now, from the definitions of the reductions and augmentations it is straightforward to check that $R(\tilde N_i,\ell)=N_i$
    and we get that $N_1\simeq N_2$, a contradiction.
  \qed
\end{pf}

We treat now the case of applying different augmentation operations to the same BTC network. But first, we give a technical lemma that will be useful in the proof of the proposition.

\begin{lem}\label{lem:automorph}
  Let $N$ be a BTC network. Then, the identity is the only automorphism (as a leaf-labeled directed graph) of $N$.
\end{lem}

\begin{pf}
    Let $\phi$ be any automorphism of $N$. Since $\phi$ is an automorphism of directed graphs and sends each leaf to itself, it follows that $\mu(u)=\mu(\phi(u))$ for each node $u$ of $N$, where $\mu(u)$ is the $\mu$-vector of $u$ as defined in \citet{Cardona2009}. Then, by \citep[Lemma~5c]{Cardona2009}, it follows that $u$ and $\phi(u)$ are either equal, or one of them is the single child of the other one; according to our definition of BTC networks, this last possibility implies that one of them is a hybrid node and the other one is a tree node, which is impossible if $\phi$ is an automorphism. Hence $\phi(u)=u$ for every node $u$.
    \qed
\end{pf}


\begin{prop}\label{thm:unicity-operation}
  Let $\tilde N_1$ and $\tilde N_2$ be two BTC networks, both obtained by one augmentation operation applied to the same BTC network $N$. If either the kinds of operation or the feasible pairs used to construct $\tilde N_1$ and $\tilde N_2$ are different, then $\tilde N_1$ and $\tilde N_2$ are not isomorphic.
\end{prop}

\begin{pf}
  Let us assume that $\tilde N_1$ and $\tilde N_2$ are isomorphic. Then, it is clear that they have the same set of labels, and exactly one of them, say $\ell$, is not a label of $N$. Since $\tilde N_1$ and $\tilde N_2$ are isomorphic, the kind of $\ell$ is the same in both networks, which implies that the kind of augmentation operations used to construct $\tilde N_1$ and $\tilde N_2$ are the same. Also, since $\tilde N_1$ and $\tilde N_2$ are isomorphic, the nodes in the respective recovering data of the reductions $R(\tilde N_i,\ell)$ must be linked by an isomorphism of phylogenetic networks. Therefore, and since by Lemma~\ref{lem:automorph} BTC networks do not have a nontrivial automorphism, the respective recovering data must be equal.
  \qed
\end{pf}

The following proposition shows that the reduction procedure defined in the previous section can be reverted using the augmentation operations presented in this section.

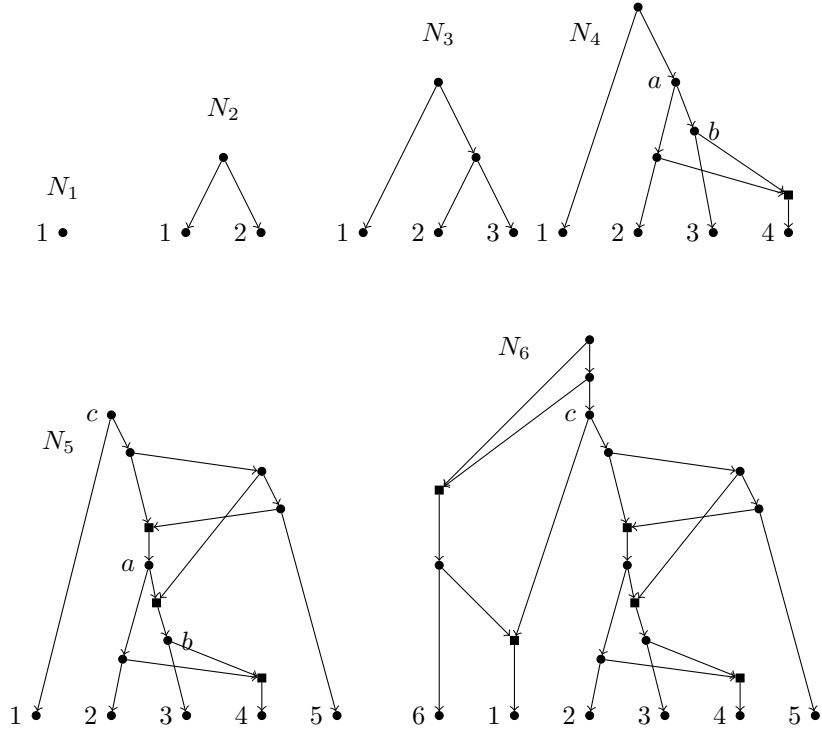
\begin{figure}
  \centering
  \begin{tikzpicture}
    \node[treenode,label=left:$1$] (l1) at (0,-7) {};
    \node[label=$N_1$] (N) at (0,-6.8) {};
\end{tikzpicture}
\qquad
 \begin{tikzpicture}
    \node[treenode] (r) at (0.5,-6) {};
    \node[treenode,label=left:$1$] (l1) at (0,-7) {};
    \node[treenode,label=left:$2$] (l2) at (1,-7) {};
    \draw[->] (r)--(l1);
    \draw[->] (r)--(l2);
    \node[label=$N_2$] (N) at (0.5,-5.75) {};
\end{tikzpicture}
 \qquad
    \vspace{0.5cm}
 \begin{tikzpicture}
    \node[treenode] (r) at (1,-5) {};
    \node[treenode] (x) at (1.5,-6) {};
    \node[treenode,label=left:$1$] (l1) at (0,-7) {};
    \node[treenode,label=left:$2$] (l2) at (1,-7) {};
    \node[treenode,label=left:$3$] (l3) at (2,-7) {};
    \draw[->] (r)--(l1);
    \draw[->] (r)--(x);
    \draw[->] (x)--(l2);
    \draw[->] (x)--(l3);
    \node[label=$N_3$] (N) at (1,-4.75) {};
\end{tikzpicture}
 \vspace{0.5cm}
 \begin{tikzpicture}
    \node[treenode] (r) at (1,-4) {};
    \node[treenode,label=left:$a$] (a) at (1.5,-5) {};
    \node[treenode] (x) at (1.25,-6) {};
    \node[treenode,label=right:$b$] (b) at (1.75,-5.65) {};
    \node[hybnode] (h) at (3,-6.5) {};
    \node[treenode,label=left:$1$] (l1) at (0,-7) {};
    \node[treenode,label=left:$2$] (l2) at (1,-7) {};
    \node[treenode,label=left:$3$] (l3) at (2,-7) {};
    \node[treenode,label=left:$4$] (l4) at (3,-7) {};
    \draw[->] (r)--(l1);
    \draw[->] (r)--(a);
    \draw[->] (a)--(x);
    \draw[->] (a)--(b);
    \draw[->] (x)--(l2);
    \draw[->] (x)--(h);
    \draw[->] (b)--(l3);
    \draw[->] (b)--(h);
    \draw[->] (h)--(l4);
    \node[label=$N_4$] (N) at (0.3,-4.75) {};
\end{tikzpicture}
  \vspace{0.5cm}
 \begin{tikzpicture}
    \node[treenode,label=left:$c$] (r) at (1,-3) {};
    \node[treenode,label=left:$a$] (a) at (1.5,-5) {};
    \node[treenode] (x) at (1.15,-6.25) {};
    \node[treenode,label=right:$b$] (b) at (1.75,-6) {};
    \node[hybnode] (h) at (3,-6.5) {};
    \node[treenode] (w) at (1.25,-3.5) {};
    \node[hybnode] (v1) at (1.6,-5.5) {};
    \node[hybnode] (v2) at (1.5,-4.5) {};
    \node[treenode] (u1) at (3,-3.75) {};
    \node[treenode] (u2) at (3.25,-4.25) {};
    \node[treenode,label=left:$1$] (l1) at (0,-7) {};
    \node[treenode,label=left:$2$] (l2) at (1,-7) {};
    \node[treenode,label=left:$3$] (l3) at (2,-7) {};
    \node[treenode,label=left:$4$] (l4) at (3,-7) {};
    \node[treenode,label=left:$5$] (l5) at (4,-7) {};
    \draw[->] (r)--(l1);
    \draw[->] (r)--(w);
    \draw[->] (w)--(u1);
    \draw[->] (w)--(v2);
    \draw[->] (u1)--(u2);
    \draw[->] (u1)--(v1);
    \draw[->] (v2)--(a);
    \draw[->] (a)--(x);
    \draw[->] (a)--(v1);
    \draw[->] (x)--(l2);
    \draw[->] (x)--(h);
    \draw[->] (v1)--(b);
    \draw[->] (b)--(l3);
    \draw[->] (b)--(h);
    \draw[->] (h)--(l4);
    \draw[->] (u2)--(v2);
    \draw[->] (u2)--(l5);
    \node[label=$N_5$] (N) at (0.3,-3.75) {};
\end{tikzpicture}
\qquad
 \begin{tikzpicture}
    \node[treenode,label=left:$c$] (r) at (1,-3) {};
    \node[treenode] (a) at (1.5,-5) {};
    \node[treenode] (x) at (1.15,-6.25) {};
    \node[treenode] (b) at (1.75,-6) {};
    \node[hybnode] (h) at (3,-6.5) {};
    \node[treenode] (w) at (1.25,-3.5) {};
    \node[hybnode] (v1) at (1.6,-5.5) {};
    \node[hybnode] (v2) at (1.5,-4.5) {};
    \node[treenode] (u1) at (3,-3.75) {};
    \node[treenode] (u2) at (3.25,-4.25) {};
    \node[hybnode] (u0) at (-1,-4) {};
    \node[treenode] (u_1) at (-1,-5) {};
    \node[treenode] (w1) at (1,-2) {};
    \node[treenode] (w2) at (1,-2.5) {};
    \node[hybnode] (v_1) at (0,-6) {};
    \node[treenode,label=left:$6$] (l6) at (-1,-7) {};
    \node[treenode,label=left:$1$] (l1) at (0,-7) {};
    \node[treenode,label=left:$2$] (l2) at (1,-7) {};
    \node[treenode,label=left:$3$] (l3) at (2,-7) {};
    \node[treenode,label=left:$4$] (l4) at (3,-7) {};
    \node[treenode,label=left:$5$] (l5) at (4,-7) {};
    \draw[->] (w1)--(w2);
    \draw[->] (w1)--(u0);
    \draw[->] (w2)--(u0);
    \draw[->] (w2)--(r);
    \draw[->] (u0)--(u_1);
    \draw[->] (u_1)--(l6);
    \draw[->] (u_1)--(v_1);
    \draw[->] (v_1)--(l1);
    \draw[->] (r)--(v_1);
    \draw[->] (r)--(w);
    \draw[->] (w)--(u1);
    \draw[->] (w)--(v2);
    \draw[->] (u1)--(u2);
    \draw[->] (u1)--(v1);
    \draw[->] (v2)--(a);
    \draw[->] (a)--(x);
    \draw[->] (a)--(v1);
    \draw[->] (x)--(l2);
    \draw[->] (x)--(h);
    \draw[->] (v1)--(b);
    \draw[->] (b)--(l3);
    \draw[->] (b)--(h);
    \draw[->] (h)--(l4);
    \draw[->] (u2)--(v2);
    \draw[->] (u2)--(l5);
    \node[label=$N_6$] (N) at (0,-2.5) {};
\end{tikzpicture}
  \caption{Example of a BTC network and the chain of augmentation operations that generate it. Namely: $N_2=T^{-1}(N_1,2;\{1\},\emptyset)$, $N_3=T^{-1}(N_2,3;\{2\},\emptyset)$,
  $N_4=H^{-1}(N_3,4;\{2,3\},\emptyset)$,
  $N_5=T^{-1}(N_4,5;\{a\},(b,a))$, and
  $N_6=H^{-1}(N_5,6;\{c,c\},(1))$.}
  \label{fig:figuraJC}
\end{figure}

\begin{prop}\label{thm:reduction-augmentation}
  Let $N$ be a BTC network and $\ell$ a leaf of $N$. 
  Let $N'=R(N,\ell)$, 
   $(S_1,S_2)$ its recovering data, and
    $\tilde N=R^{-1}(N',\ell,S_1,S_2)$. 
    Then, $N$ and $\tilde N$ are isomorphic.
\end{prop}
\begin{pf}
  It is straightforward to see that the operations $T^{-1}$ and $H^{-1}$ reverse the effects of $T$ and $H$, respectively. The only points worthy of attention correspond to the cases where the single node in $S_1$ appears in $S_2$ (for reductions/augmentations of type $T$) or where there is a single node in $S_1$ with multiplicity two (for reductions/augmentations of type $H$). In the first case, the augmentation process creates two elementary nodes, $w_1$ and $v_i$, connected by an arc $(w_1,v_i)$, which is the same situation as in $N$ after the removal of the nodes in $\OTHP(\ell)$. In the second case, two elementary nodes $\tau_1$ and $\tau_2$ are created, connected by an arc, once again the same situation as in $N$ after the removal of the nodes in $\OTHP(\ell)$.
\qed
\end{pf}
A direct consequence of  the results in this section is the following theorem, which can be used to generate in an effective way all BTC networks over a set of taxa. See Figure~\ref{fig:figuraJC} for an example.
%
%
\begin{thm}
  Let $N\in\BTC_n$ be a BTC network over $[n]$. Then, $N$ can be constructed from the trivial network in $\BTC_1$ (with one node labeled by $1$) by application of $n-1$ augmentation operations, where at each step $i$, the leaf $i+1$ is added. Moreover, these augmentation operations are unique.
\end{thm}
\begin{pf}
  The existence is a direct consequence of Corollary~\ref{cor:decomposition-of-network} and Proposition~\ref{thm:reduction-augmentation}. Unicity comes from
  Propositions~\ref{prp:different_nets} and~\ref{thm:unicity-operation}.
\qed
\end{pf}

It should be noted that very recently, other methods to generate all BTC networks over a set of taxa have been proposed \citep{orchard}, but,  to our knowledge, this is the first time that the networks are generated with unicity. In previous attempts, an 
isomorphism check was needed after the generation phase. 
\section{Bounding the number of networks}
\label{sec:bounding-offspring}

In this section, we shall first give bounds for the number of BTC networks that can be obtained from a given one by means of augmentation operations. This will be done by bounding the number of feasible pairs in such a network. Then, we shall find bounds for the number of BTC networks with a fixed number $n$ of leaves.


Let $N$ be a BTC network over $[n]$ with $h$ hybrid nodes. From Lemma~\ref{lem:number_nodes} we know that it has $t=2n+h-1$ tree nodes, and that $h\le n-1$ and $t\le 3n-2$.
%
In the following, we shall show how to 
compute the number of pairs $(S_1,S_2)$ satisfying all conditions of Definition \ref{dfn:feasible}, except for Condition 3, via an auxiliary problem. Note that this will only give an upper bound for the number of networks, since the pairs we find can  produce networks with cycles. 

%
%

\subsection{An auxiliary problem}

Let $P(N,k)$ be the set of tuples of length $k$ of tree nodes of $N$ such that (1) no pair of them are equal or siblings, and (2) none of them has a hybrid parent or sibling. We indicate the number of such tuples as $p(N,k)=|P(N,k)|$, and since this number will only depend on $n$, $h$ and $k$, we indicate it also by $p(n,h,k)$. We consider the problem of computing $p(n,h,k)$.


We compute first how many tree nodes are there that do not have neither a hybrid parent nor a hybrid sibling. Since the single child of a hybrid node must be a tree node, there are $h$ tree nodes that have a hybrid parent. Note that each hybrid node has two siblings that must be tree nodes; also, a tree node cannot be sibling of two different hybrid nodes; hence, there are $2h$ tree nodes that have a hybrid sibling. Since there cannot be a tree node having the two properties (if it has a hybrid parent, then it does not have any hybrid sibling), there are $3h$ tree nodes that are either a child or a sibling of a hybrid node.
Then, the number of tree nodes that do not have neither a hybrid parent nor a hybrid sibling is $t-3h=2n-2h-1$. Note that this set of nodes is composed by the root of the network and pairs of tree nodes that are siblings.

Consider now the problem of counting the number of tuples $(y_1,\dots,y_k)$ in this set that are neither equal nor siblings. We distinguish two cases:
\begin{itemize}
\item If none of the nodes $y_i$ is the root of $N$, we start having $2n-2h-2$ choices for $y_1$, and at each stage the number of choices decreases of two units. Hence, the number of choices is
  $$
\begin{aligned}
    p_0(n&,h,k)=\\
    &=(2n-2h-2)(2n-2h-4)\cdots(2n-2h-2k)\\
    &=2^k(n-h-1)(n-h-2)\cdots(n-h-k)\\
    &=2^k\frac{(n-h-1)!}{(n-h-k-1)!}.
\end{aligned}
$$
\item If one of the nodes $y_i$ is the root of $N$, then the process of constructing an element in $P(N,k)$ can be described as first choosing at which position $i$ one puts the root, and then filling in the remaining $k-1$ positions with 
a tuple of the set $P(N,k-1)$ such that none of the nodes is the root (which is what we have just computed). Hence, the number of possibilities is
  $$p_1(n,h,k)=k 2^{k-1}\frac{(n-h-1)!}{(n-h-k)!}.$$
\end{itemize}
Then we get that
$$
\begin{aligned}
p(n,h,k)&=p_0(n,h,k)+p_1(n,h,k)\\
&=2^k\frac{(n-h-1)!}{(n-h-k-1)!}+k 2^{k-1}\frac{(n-h-1)!}{(n-h-k)!}
\end{aligned}
$$
\subsection{Counting pairs satisfying Conditions 1, 2 and 4H}
\label{sec:counting-h-feasible}
Consider pairs $(S_1,S_2)$ satisfying Conditions 1, 2 and 4H. 
Recall that, since condition 4H implies that $S_1$ and $S_2$ cannot have elements in common, Conditions 1 and 2 are simplified: no pair of nodes in $S_2$ can be siblings and none of them can either be the child of a hybrid node or have a hybrid sibling. Hence,
the problem is equivalent to finding a tuple $(y_1,\dots,y_{r-1})$ in $P(N,r-1)$ and then either a tree node $\tau_1$ or an unordered pair $\{\tau_1,\tau_2\}$ of different tree nodes, in either case disjoint from those in $(y_1,\dots,y_{r-1})$. Once the tuple $(y_1,\dots,y_{r-1})$ is fixed, the number of tree nodes available for choosing $\tau_1$ and $\tau_2$ is $t-r+1=2n+h-r$.
Hence, the number of possible pairs is $$F_H(n,h,r-1)=F_{H,1}(n,h,r-1)+F_{H,2}(n,h,r-1),$$ 
where
$$
\begin{aligned}
F_{H,1}(n,h,r-1)&=p(n,h,r-1)\cdot (2n+h-r),\\
F_{H,2}(n,h,r-1)&=p(n,h,r-1)\cdot{}\\
&\qquad\cdot\frac12(2n+h-r)(2n+h-r-1).
\end{aligned}
$$

\subsection{Counting pairs satisfying Conditions 1, 2 and 4T}
The problem now is to count the ways of choosing $S_1=\{\tau_1\}$ and a tuple  $S_2=(y_1,\dots,y_{r-1})$ satisfying Conditions 1, 2 and 4T. Now $\tau_1$ can appear in $S_2$, and different possibilities arise, since it allows that one of the nodes in $S_2$ has a sibling in $S_2$, or that it has a hybrid parent or sibling. We consider, thus, these different possibilities:
\begin{itemize}
\item $\tau_1\neq y_i$ (for all $i$): This case is very similar to one considered in the previous paragraph, specifically the case where only a single node $\tau_1$ had to be taken. The number of possible pairs is
$$F_{T,1}(n,h,r-1)=p(n,h,r-1)\cdot(2n+h-r).$$
\item $\tau_1=y_i$ is a child or a sibling of a hybrid node: Choosing one of these pairs is equivalent to first choosing the position $i$, then filling the other $r-2$ positions with a tuple in $P(N,r-2)$, and then choosing a node that is a child or sibling of a hybrid node to be put in the position $i$. The number of ways to do this procedure is
  $$F_{T,2}(n,h,r-1)=p(n,h,r-2)\cdot (r-1)\cdot 3h,$$
  since each hybrid node has a single child and two siblings, and none of these $3h$ nodes appears twice, associated to two different hybrid nodes.
\item $\tau_1=y_i$ is a sibling of some other node $y_j$ in $S_2$: In this case one has to choose the positions $i$ and $j$ where to put the pair of sibling nodes, fill the other $r-3$ positions with a tuple in $P(N,r-3)$, choose a pair of sibling tree nodes to take as $y_i$ and $y_j$, and finally set $\tau_1=y_i$. The choice of $i$ and $j$ can be done in $(r-1)(r-2)$ different ways. The choice of the tuple of length $r-3$ can be done in $p(n,h,r-3)$ ways; $p_1(n,h,r-3)$ of them contain the root of $N$ (and $r-4$ tree nodes with a sibling tree node) and $p_0(n,h,r-3)$ do not contain the root (and contain $r-3$ tree nodes with a sibling tree node). Once this is done, the number of available pairs of sibling tree nodes is $n-h-1-(r-4)=n-h-r+3$, if the root of $N$ was chosen, or $n-h-1-(r-3)=n-h-r+2$ otherwise. 
  Hence, the total number of pairs is $F_{T,3}(n,h,r-1)=F_{T,3,A}(n,h,r-1)+F_{T,3,B}(n,h,r-1)$, corresponding to these two cases, with:
    $$
  \begin{aligned}
  F_{T,3,A}(n,h,r-1)&=(r-1)(r-2)\cdot p_1(n,h,r-3)\cdot{}\\
  &\qquad\cdot(2n-2h-2r+6),\\
  F_{T,3,B}(n,h,r-1)&=(r-1)(r-2)\cdot p_0(n,h,r-3)\cdot{}\\
  &\qquad\cdot(2n-2h-2r+4).
\end{aligned}
  $$
  
\item $\tau_1=y_i$ but none of the previous conditions hold: In this case one only has to take a tuple in $P(N,r-1)$ and choose which of the $r-1$ nodes to take as $\tau_1$. The number of possible pairs is then
  $$F_{T,4}(n,h,r-1)= p(n,h,r-1)\cdot (r-1).$$
\end{itemize}
Note that the four conditions above are mutually exclusive. Hence, the overall number of possible pairs $(S_1,S_2)$ is the sum of all numbers found:
$$
  \begin{aligned}
F_T(n,h,r-1)&=F_{T,1}(n,h,r-1)+F_{T,2}(n,h,r-1)+{}\\
&\qquad+F_{T,3}(n,h,r-1)+F_{T,4}(n,h,r-1). 
\end{aligned}
$$

\subsection{Bounds for the number of networks}


Each network $N\in\BTC_n$ with $h$ hybrid nodes, appears as augmentation $R^{-1}(N',n,S_1,S_2)$
of an unique network $N'\in\BTC_{n-1}$ with $h'$ hybrid nodes, where $S_2$ has length $r-1=h-h'$, if the augmentation is of type $T$, or $r-1=h-h'-1$ if it is of type $H$. If we call $B(n,h)$ the number of networks in $\BTC_n$ with $h$ hybrid nodes, and since we have bounded the number of feasible pairs, we have that
$$\begin{aligned}
B(n,h)&\le
\sum_{h'=0}^h B(n-1,h')\cdot F_T(n-1,h',h-h')+{}\\
&\qquad +
\sum_{h'=0}^{h-1} B(n-1,h')\cdot F_H(n-1,h',h-h'-1)
\end{aligned}
$$

Also, since the number of hybrid nodes in a BTC network with $n$ leaves is at most $n-1$, we have that
$$|\BTC_n|=\sum_{h=0}^{n-1}B(n,h),$$
and the expression above allows us to compute a bound for this number of networks. See section~\ref{sec:computations} for an experiment with these bounds.

The asymptotic formula $|\BTC_n|= 2^{2n \log n+ O(n)}$ is given in \citet{McDiarmid2015}, and both our experimental results in Section~\ref{sec:computations} for $n\le 7$ and the bounds that we have computed for $n\le 10$ 
are coherent with this expression.
However, the problem of finding a closed expression for the asymptotic behaviour of our bounds is still open.


\section{Computational experiments}
\label{sec:computations}

The algorithms in this paper have been implemented in python using the python library PhyloNetworks \citep{pypi_phylonetworks}. This implementation, together with the sources for the experiments that we comment in this section can be downloaded from \url{https://github.com/bielcardona/TCGenerators}. 

\paragraph{Exhaustive and sequential construction of networks in $\BTC_n$.}
We have implemented  both the exhaustive and sequential construction of BTC networks with $n$ leaves. 
The number of such networks increases very rapidly, and hence the exhaustive construction is not feasible for $n\ge 8$. For $n\le 6$ we generated all the networks in $\BTC_n$; see Table~\ref{number-btc} for the number of such networks. For $n=7$ we could not find in a reasonable time all the networks (in our implementation on a cluster of 32 CPUs it would have taken 15 days). Instead, we took uniform samples of networks in $\BTC_6$ and computed their respective offspring, and repeated this procedure until the average number of offsprings per network stabilized up to 4 digits; this allowed us to give an estimate for $|\BTC_7|$.

\paragraph{Random construction of networks in $\BTC_n$.}
We have implemented the following construction, that does not generates networks uniformly, but is the closest we could get to it. We start with the network $N_1$ with a single node labeled by $1$. At each stage $i=1,\dots,n-1$, we explicitly find all feasible pairs inside $N_i$ and choose at random and uniformly one of them to generate the network $N_{i+1}$. This procedure generates all possible networks in $\BTC_n$, but not uniformly, since different networks over the same set of taxa may have different number of feasible pairs.

\paragraph{Computation of bounds for $|\BTC_n|$.}
Finally, we have implemented the recursive computation for the upper bounds of $|\BTC_n|$ using the bounds for the offsprings of BTC networks found in Section~\ref{sec:bounding-offspring}. The results for $n$ up to $10$ are given in Table~\ref{number-btc}, where it is observed that, at least for small values of $n$, the true number of networks and the upper bounds have the same order of magnitude.

\begin{table}
  \centering
  \begin{tabular}{rrr}
    \toprule
    $n$&$|\BTC_n|$& upper bound\\\midrule
1 & 1 & 1 \\
2 & 3 & 3 \\
3 & 66 & 85 \\
4 & 4,059& 7,442 \\
5 & 496,710& 1,317,098 \\
6 &             101,833,875 & 387,405,870 \\
7 & $\simeq$ 31,500,000,000 & 169,781,857,790 \\
8 & ? & 103,409,407,515,286 \\
9 & ? & 83,400,205,845,281,275 \\
10 & ? & 85,947,517,732,640,544,027 \\
    \bottomrule
  \end{tabular}
  \caption{Exact number of BTC networks over $[n]$ for $n=1,\dots,6$, an estimate for $n=7$, and their upper bounds for $n\le10$. }
  \label{number-btc}
\end{table}

\section{Conclusion\label{sec:conclusions}}

The main result of this paper is a systematic way of recursively generating, with unicity, all BTC networks with a given number of leaves. This procedure relies on a pair of reduction/augmentation operations that generalize  analogous operations for phylogenetic trees. 
Indeed, given a (rooted, binary) phylogenetic tree over $[n]$, we can obtain a phylogenetic tree over $[n-1]$  by deleting the leaf labeled by $n$ and removing the elementary node that this deletion generates. Conversely, given a tree $T$ over $[n-1]$ and one of its nodes $u$, we can construct a tree over $[n]$ by simply hanging a pendant leaf labeled by $n$ to the single incoming arc of $u$. Since different choices for $T$ and $u$ give different trees over $[n]$, this gives  a recursive procedure to generate, with unicity, all binary rooted phylogenetic trees over a given set of taxa: we start with the leaf labeled by $1$, then we add the leaf labeled by $2$, then the leaf labeled by $3$ in all possible ways, and so on.
Biologically, we can think of this procedure as follows: Once the evolutionary history of a given set of OTUs is correctly established\footnote{In practice, we can never be sure that we got the correct tree, but here we suppose we do.} and  modeled by a phylogenetic tree, extending this evolutionary history  to consider a `new''  OTU $n$ consists in finding where to place $n$ in the tree, i.e. finding the speciation event that leads to the diversification of $n$.

Unfortunately, when working with classes of phylogenetic networks, the removal of a single leaf (and of all elementary nodes created by this removal) does not necessarily give a phylogenetic network within the same class. In the case of BTC networks, we were able to find the minimal set of nodes that one must remove so that, after their deletion and that of all elementary nodes created by this removal, one gets a BTC network with one leaf less. 
As in the case of trees, given a BTC network over $[n-1]$ and some set of nodes with certain restrictions (i.e. the feasible pairs $S_1$ and $S_2$) we can construct a BTC network over $[n]$ leaves, in such a way that different choices for the BTC network or for the feasible pair give different BTC networks over $[n]$. Hence, we find a procedure to recursively generate all BTC networks over a given set of taxa.
Biologically, we can think of this procedure as an extension of what can happen when adding a new OTU $n$ to a phylogenetic tree: here the diversification of $n$ can involve a reticulated event (when $n$ is added as hybrid node) and  the ancestors of $n$  participate to  $|S_2|$ reticulated events, which were 
impossible to detect before the introduction of $n$. 



\bibliographystyle{plainnat}
\bibliography{biblio}{}

\end{document}